\documentclass[twocolumn,floatfix,showpacs,preprintnumbers,amsmath,amssymb,prb]{revtex4}
\usepackage{graphicx}
\usepackage{dcolumn}
\usepackage{bm}
\usepackage[dvips]{color}

\begin{document}

\title{Carrier-number fluctuations in the 2-dimensional electron gas at the LaAlO$_{3}$/SrTiO$_{3}$ interface}

\author{C. Barone$^{1}$}
 \email{cbarone@unisa.it}
\author{F. Romeo$^{1}$, S. Pagano$^{1}$, E. Di Gennaro$^{2}$, F. Miletto Granozio$^{2}$, I. Pallecchi$^{3}$, D. Marr\`e$^{3}$}
\author{U. Scotti di Uccio$^{2}$}

\affiliation{$^{1}$Dipartimento di Fisica "E.R. Caianiello" and CNR-SPIN Salerno, Universit\`a di Salerno, I-84084 Fisciano, Salerno, Italy\\$^{2}$CNR-SPIN Napoli and Dipartimento di Fisica, Universit\`a di Napoli "Federico II", I-80126 Napoli, Italy\\$^{3}$CNR-SPIN Genova and Dipartimento di Fisica, Universit\`a di Genova, I-16152 Genova, Italy}


\begin{abstract}
The voltage-spectral density $S_{V}(f)$ of the 2-dimensional electron gas formed at the interface of LaAlO$_{3}$/SrTiO$_{3}$ has been thoroughly investigated. The low-frequency component has a clear 1/f behavior with a quadratic bias current dependence, attributed to resistance fluctuations. However, its temperature dependence is inconsistent with the classical Hooge model, based on carrier-mobility fluctuations. The experimental results are, instead, explained in terms of carrier-number fluctuations, due to an excitation-trapping mechanism of the 2-dimensional electron gas.
\end{abstract}


\maketitle

Polar/non-polar oxide interfaces can host a 2-dimensional electron gas (2DEG).\cite{Ohtomo} A well-known example is the LaAlO$_{3}$/SrTiO$_{3}$ interface (LAO/STO), made by depositing an epitaxial LAO film onto a single-crystal STO with single TiO$_{2}$-plane termination. LAO/STO interfaces are conducting, provided LAO thickness exceeds the threshold of $4$ unit cells.\cite{Thiel} The properties of LAO/STO heterojunctions have been widely investigated (see, e.g., Refs.~\onlinecite{Mannhart} and \onlinecite{Zubko}), and it is known that the mobile electrons populate a quantum well, within the STO side of the junction, with a confinement depth of few nanometers.\cite{Copie,Popovic,Cantoni} The conductance band DOS is constituted of three overlapped 3d t$_{2g}$ sub-bands (namely, Ti3d$_{xy}$, Ti3d$_{xz}$, and Ti3d$_{yz}$), with lifted degeneracy due to the breakdown of lattice symmetry imposed by the interface, and populated by charges injected from the donor states. The nature of such donors is still controversial. They are mainly identified either as valence band or in-gap states located at the LAO surface (electronic reconstruction model),\cite{Popovic,Santander,Berner} or as defect states located in STO (models based on cation intermixing or oxygen vacancies).\cite{Chambers} The electronic conduction can be qualitatively understood in terms of charge carrier of the 2DEG hopping between otherwise empty Ti$^{4+}$ (i.e., Ti(d$^{0}$)) sites. Such states are prone to localization, in the presence of either interface disorder, due to their bidimensional nature (Anderson localization), or of a strong coupling with the lattice (polaronic localization),\cite{Devreese} that might well be attributed to the Jahn-Teller effect taking place in Ti(d$^{1}$) cations. It has been proposed in Ref.~\onlinecite{Popovic} that such localization mainly affects the Ti3d states deriving from the interfacial TiO$_{2}$ layer, and in particular the low-energy Ti3d$_{xy}$ sub-band, while Ti3d states from the very next atomic layers might host mobile electrons.

LAO/STO shows a metallic sheet resistance vs. temperature $R_{\Box}(T)$ in a wide range of temperature, with typical value $R_{\Box} \geq 10$~k$\Omega$/$\Box$ at room temperature, that drops to zero at the superconducting transition, normally in the mK range.\cite{Caviglia} The carrier density and mobility are usually determined from Hall effect and magnetoresistance measurements, including the observation of Schubnikov-De Haas oscillations.\cite{Caviglia2} In particular, the 2DEG charge carrier density ranges from $10^{13}$ to $10^{14}$~cm$^{-2}$, depending on the fabrication conditions, with a sensitive variation on temperature (see, e.g., Ref.~\onlinecite{Dubroka}). The mobility is also affected by the fabrication process; typical values are about $\approx 10$~cm$^{2}$V$^{-1}$$s^{-1}$ at room temperature and reach up to $\approx 10^{3}$~cm$^{2}$V$^{-1}$$s^{-1}$ at low temperature. As a general phenomenological trend, the lower is the charge carrier density, the higher the mobility.\cite{Huijben}

\begin{figure}[!b]
\resizebox{1.0\columnwidth}{!}{%
  \includegraphics{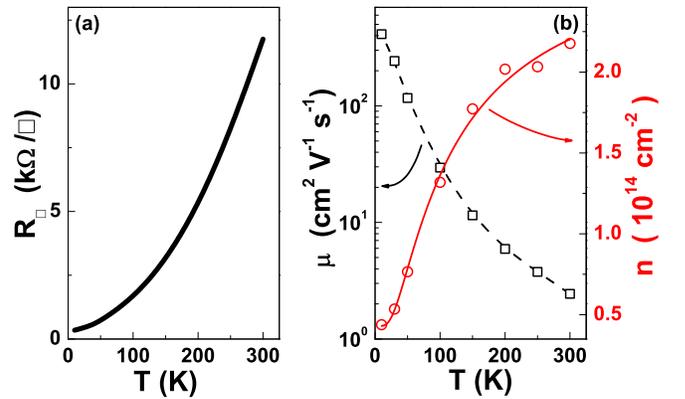}
} \caption{\label{fig:Fig1} (Color online) (a) $R_{\Box}(T)$; (b) $\mu(T)$ (left axis) and $n(T)$ (right axis). The solid line is a fit (see Eq.~(\ref{eq:nvsT}) in the text), the dashed line is a guide to the eye.}
\end{figure}

\begin{figure*}[!t]
\resizebox{1.6\columnwidth}{!}{%
  \includegraphics{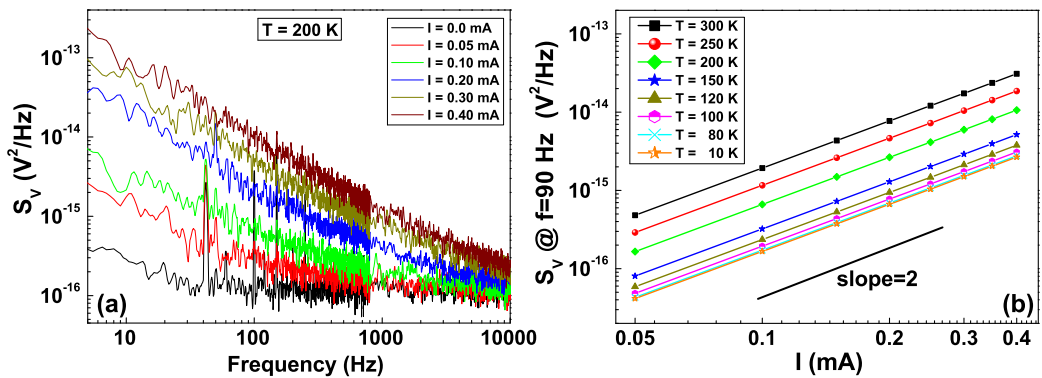}
} \caption{\label{fig:Fig2} (Color online) (a) Voltage spectral traces at $200$~K and for different bias currents. A change in data density, due to the choice of different frequency spans, is evident at $f=800$~Hz. (b) Current dependence of the voltage-spectral density at $90$~Hz, in the temperature range $10-300$~K. The solid line with slope $2$ in the log-log graph is a reference for the quadratic scaling.}
\end{figure*}

In view of the possible applications of LAO/STO based devices, it is important to characterize the transport properties by also addressing the intrinsic electric noise. Noise in semiconductors has been deeply investigated;\cite{2DEG-Semic} while, besides a few seminal works on the subject, the electric noise in conducting oxides remains a young and stimulating field of research, still open to both theoretical and experimental activity.\cite{MangaNoise} The so-called "electric noise spectroscopy" has already demonstrated good potential for the comprehension of non conventional transport phenomena in oxides (see, e.g., Refs.~\onlinecite{Barone-LSMO,Barone-PCMO,Barone-WL}). It may, therefore, contribute to clarify still unsolved fundamental questions regarding the electronic properties of LAO/STO.

A detailed characterization and interpretation of the noise spectral density in LAO/STO interface is here reported. The samples were fabricated by RHEED-assisted laser ablation, following standard procedures described in Ref.~\onlinecite{Napoli-Interfaces}, giving sharp, high quality structures,\cite{Cantoni} and reproducible transport properties.\cite{DiGennaro} Au wires were bonded to the samples by ultrasonic bonder in the standard four-probe Van der Pauw geometry. $R_{\Box}(T)$ and Hall effect measurements were carried out on a batch of samples in a He-flux cryostat from $2.5$~K to room temperature and in magnetic fields up to $9$~T. Resistance and I-V curves were measured both in constant and pulsed mode. Current values from $0.05$ to $0.45$~mA were used, in order to rule out Joule-heating effects. Noise measurements were made in a closed-cycle refrigerator at temperatures between $8$ and $300$~K, with a stability better than $0.1$~K. A low-noise digital current supply was used for biasing the samples. The output voltage signal, amplified with a low-noise PAR5113 preamplifier, was acquired by dynamic signal analyzer HP35670A. The equivalent input voltage-spectral density, due to the electronic chain, was typically $S_{V_{n}} \mathcal{\simeq} 1.4 \times 10^{-17}$~V$^{2}$/Hz. A specific experimental procedure, described in detail in Ref.~\onlinecite{Barone-NoiseRed}, was applied to minimize the electrical noise generated by the contacts.

In the following the data from a representative LAO/STO sample with $12$ unit cells thick LAO film grown on STO substrate at $700$~$^\circ$C in an oxygen pressure $P(O_{2})=10^{-2}$~mbar are reported. The sheet resistance $R_{\Box}(T)$ is shown in Fig.~\ref{fig:Fig1}(a); the carrier density $n(T)$ and the mobility $\mu(T)$ (as deduced by Hall effect measurements in a single band approximation) are shown in Fig.~\ref{fig:Fig1}(b). The temperature dependence of the carrier density can be satisfactorily described by (see the solid line in Fig.~\ref{fig:Fig1}(b))
\begin{equation}
n\left(T\right)=n_{0}+n_{1}exp\left(-\frac{E_{a}}{k_{B}T}\right) \label{eq:nvsT}~,
\end{equation}
where $n_{0}$ and $n_{1}$ are mobile and trapped charge carriers, respectively. The best fitting parameters are $n_{0}=(0.43 \pm 0.05) \times 10^{14}$~cm$^{-2}$, $n_{1}=(2.44 \pm 0.09) \times 10^{14}$~cm$^{-2}$, $E_{a}=(8.3 \pm 0.3)$~meV. From Eq.~(\ref{eq:nvsT}), at $T=0$ the conduction band is populated with $n_{0}$ charge carriers, while $n_{1}$ are trapped in a very narrow band at energy $E_{a}$ below the chemical potential $\mu^{*}$. These trapped states may be interpreted as large polarons or as Anderson-localized states. The thermal activation of localized electrons promotes a consistent number of carriers to the mobile states, and explains the strong temperature dependence of $n\left(T\right)$ seen by the Hall effect measurements. In this context, the total number of electrons in the quantum well is given by $n_{0}+n_{1}$.

\begin{figure}[!t]
\resizebox{0.75\columnwidth}{!}{%
  \includegraphics{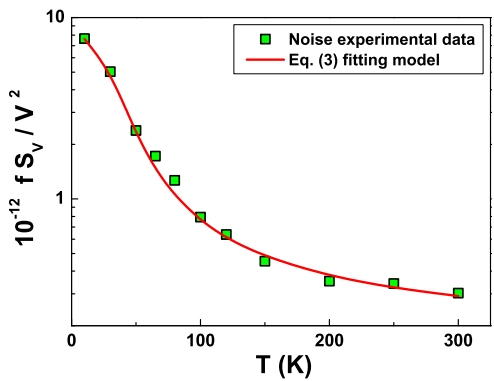}
} \caption{\label{fig:Fig3} (Color online) Normalized noise level $fS_{V}/V^{2}$ vs. $T$ (dots). Red solid curve is the fitting function given in Eq.~(\ref{eq:NoiseLevel}) (see text for the details).}
\end{figure}

In Fig.~\ref{fig:Fig2}(a) the spectral noise $S_{V}(f)$ measured at $T=200$~K is shown for different bias currents. At all temperatures, the frequency dependence of the spectral traces is characterized by a pure 1/f behavior (Flicker noise)\cite{1/fNote} at low frequencies, while, at higher frequency, $S_{V}(f)$ tends to a constant value. The latter being the sum of the Johnson noise $4k_{B}TR(T)$, and the background readout electronic noise. The sharp peaks in the spectra are a spurious effect, due to coupling to external noise sources and are excluded from the noise analysis. The dependence of $S_{V}(f)$ on the bias current is shown in Fig.~\ref{fig:Fig2}(b) and is quadratic in the whole temperature range, with an exponent $2.0 \pm 0.1$. This is a clear indication that the 1/f noise is generated by resistance fluctuations.\cite{Kogan} It is, therefore, possible to define a "normalized noise level", i.e., the product $fS_{V}(f)/V^{2}$, that is a function of the temperature only, as shown in Fig.~\ref{fig:Fig3}.

By standard statistics it is obtained
\begin{equation}
\frac{\delta V^{2}}{V^{2}}=\frac{\delta R^{2}}{R^{2}}=\frac{\delta N^{2}}{N^{2}}+\frac{\delta \mu^{2}}{\mu^{2}} \label{eq:fluctuations}~,
\end{equation}
where $\delta R^{2}$, $\delta N^{2}$, and $\delta \mu^{2}$ are the variances of the sample resistance $R$, of the total carrier number $N$ (with $N=An$, $A$ being the effective sample area), and of the electron mobility $\mu$, respectively. The relation in Eq.~(\ref{eq:fluctuations}) highlights two possible sources of resistance fluctuations, i.e., carrier-number fluctuations and mobility fluctuations. The latter is always dominating in metals. In the high-frequency range, $\delta \mu^{2}$ is mainly determined by the thermal fluctuations of the mean free path, leading to the Johnson white noise. At low frequencies, the statistical variance of the mean free path dominates. The mobility is connected to the ensemble average of the mean free path; therefore, its variance follows the scaling law $\delta \mu^{2} \propto N^{-1}$. The phenomenological Hooge formula captures this concept; the Flicker noise level due to mobility fluctuations is then given by $fS_{V}(f)/V^{2}=\alpha_{H}N^{-1}$, with $\alpha_{H} \approx 2 \times 10^{-3}$.\cite{Kogan} The results for LAO/STO do not follow this behavior. The data, instead, are very well fitted by
\begin{equation}
f\frac{S_{V}(f)}{V^{2}}=\frac{K}{2A^{2}n(T)^{2}} \label{eq:NoiseLevel}~,
\end{equation}
where $n\left(T\right)$ is the total carrier density given by Eq.~(\ref{eq:nvsT}). This implies that the noise level scales with the inverse \emph{square} of the mobile carrier number. The red solid curve in Fig.~\ref{fig:Fig3} is the plot of the function in Eq.~(\ref{eq:NoiseLevel}), with $K/2A^{2}=(1.41 \pm 0.04) \times 10^{16}$~cm$^{-4}$. In consideration of Eq.~(\ref{eq:fluctuations}), these results on LAO/STO are interpreted as indicative of carrier-number fluctuations.

A similar behavior has been described by McWhorter,\cite{McWhorter} and observed in semiconducting structures hosting a 2DEG.\cite{Noise-Semic} In this work, the model of carrier-number fluctuations has been developed for LAO/STO 2DEG to explain both the 1/f dependence, and the $1/n^{2}$ dependence of $S_{V}(f)$. The full details of the model derivation are reported as Supplemental Material (see Ref.~\onlinecite{SuppMat}). The main assumptions are given in the following, together with the outline of the mathematical procedure.

\begin{figure}[!b]
\resizebox{0.7\columnwidth}{!}{%
  \includegraphics[clip]{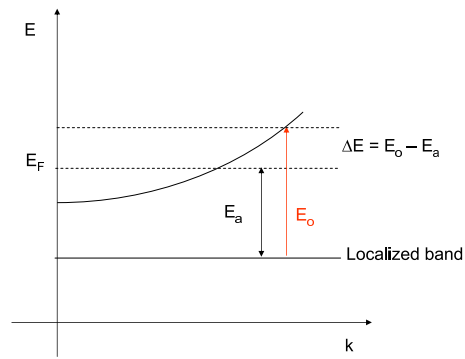}
} \caption{\label{fig:Fig4} (Color online) Sketch of a typical activation mechanism of 2DEG system at oxide interfaces. The vertical transitions, occurring between the flat localized band and unoccupied states above the Fermi level, require an activation energy $E_{0}>E_{a}$ (see main text for details). The dissipation energy $\Delta E=E_{0}-E_{a}$ is related to a phonon emission.}
\end{figure}

Referring to Eq.~(\ref{eq:nvsT}), due to the stochastic processes of excitation-trapping, the number of mobile electrons deviates from its equilibrium value. This process is described by the function $\Delta N(t)$, that satisfies the rate equation
\begin{equation}
\frac{d}{dt}\Delta N=-\frac{\Delta N}{\tau}+\xi(t) \label{eq:RateEq}~,
\end{equation}
where $\xi(t)$ is an uncorrelated noise generator, and $\tau$ is the characteristic time for the equilibration of mobile electrons with traps. After some manipulations,\cite{SuppMat} from Eq.~(\ref{eq:RateEq}) it is possible to obtain the Fourier transform of $\Delta N$, $\Delta \widetilde{N}(\omega)$, and the spectral noise density as
\begin{equation}
S_{N}(\omega)=\left\langle\left|\Delta \widetilde{N}(\omega)\right|^{2}\right\rangle=\frac{2\sigma^{2}\tau}{1+\omega^{2}\tau^{2}}=\frac{2\sigma^{2}\tau}{1+\left(2\pi\tau\right)^{2}f^{2}} \label{eq:LorentzianNoise}~,
\end{equation}
where $\omega=2\pi f$, and $\sigma^{2}$ is the variance on the stochastic variable $N$. In order to determine $\tau$, the mechanism of excitation-trapping in oxides has to be considered. It is assumed that the excitation from a trap is mainly due to scattering with acoustic phonons, which carry negligible moment (i.e., it is vertical in the k-space). According to the sketch in Fig.~\ref{fig:Fig4}, such process requires an activation energy $E_{0}$, which is supposed to be uniformly distributed above the threshold $E_{a}$. Therefore, when substituting $\tau=\tau_{0}\exp\left(E_{0}/k_{B}T\right)$ in Eq.~(\ref{eq:LorentzianNoise}) and summing over $E_{0}$, it is obtained: $S_{N}(\omega)=\sigma^{2}/2f$. Recalling that $S_{N}/N^{2}=S_{V}/V^{2}$, the normalized voltage-spectral density is derived as
\begin{equation}
f\frac{S_{V}(f)}{V^{2}}=\frac{\sigma^{2}}{2A^{2}n\left(T\right)^{2}} \label{eq:FinalEq}~.
\end{equation}
Equation~(\ref{eq:FinalEq}) is the same expression as Eq.~(\ref{eq:NoiseLevel}), where $K$ is identified with the temperature-independent variance $\sigma^{2}$. A more refined calculation of $\sigma^{2}$, reported in Ref.~\onlinecite{SuppMat}, shows that indeed its temperature dependence is very weak. As a final comment, the noise behavior of the LAO/STO samples studied here is significantly different from that of doped STO.\cite{Guerrero} In the latter case the noise level does not scale with the inverse square of carriers number. However, doped STO shows a very high mobility, indicating the suppression of electron self-trapping; hence, it is reasonable to infer that the noise generation in STO depends on mobility fluctuations rather than on carrier-number fluctuations.

In conclusion, transport properties of LAO/STO interface have been characterized in detail by Hall effect and spectral noise $S_{V}(f)$ measurements. The experimental findings support the existence of a thermally activated mechanism which promotes charge carriers from a narrow, underlying band (attributed to polarons or to Anderson-localized states), to the conduction band. The low-frequency Flicker noise spectra have pure 1/f dependence, with a $I^{2}$ scaling, allowing the attribution of the noise to resistance fluctuations. The noise level dependence on temperature cannot be consistently explained in terms of the classical Hooge picture. It is, instead, shown that a model based on the fluctuation of the carrier number describes correctly the experimental observations.


\begin{thebibliography}{99}
\bibitem{1/fNote}
By fitting the low-frequency regions with the function $S_{V}(f)=\frac{K}{f^{\alpha}}$, the best fit value $\alpha=(1.0 \pm 0.1)$ was obtained.
\end{thebibliography}
\end{document}